\documentclass[amsmath,amssymb]{revtex4}
\usepackage{graphicx}% Include figure files

\begin{document}

\title{\boldmath Determination of the branching fractions for 
$\psi(3770) \rightarrow D^0 \bar D^0, D^+D^-,D\bar D$ and
$\psi(3770) \rightarrow {\rm non-}D\bar D$}

\author{Gang RONG~~~~Dahua ZHANG~~~~Jiangchuan CHEN}

\vspace{16mm}

\affiliation{Institute of High Energy Physics, Beijing 100039, People's Republic of
China}
\vspace{18mm}

\begin{abstract}
  The branching fractions for 
$\psi(3770) \rightarrow D^0 \bar D^0, D^+D^-,D\bar D$
and $\psi(3770) \rightarrow {\rm non-}D\bar D$ are determined based on
the cross sections for $\psi(3770)$ and $D \bar D$ production
measured by BES Collaboration.
From its recent publications we determine the branching fractions for
$\psi(3770) \rightarrow D^0 \bar D^0, D^+D^-$
and $\psi(3770) \rightarrow D\bar D$ to be
$(52.2\pm 4.8\pm 5.5)\%$, $(37.0 \pm 5.0 \pm 4.3)\%$
and $(89.1 \pm 6.9 \pm 9.2)\%$, respectively.
The latter one implies the branching fraction for
$\psi(3770) \rightarrow {\rm non-}D\bar D$ to be
$(10.9 \pm 6.9 \pm 9.2)\%$,
corresponding the inclusive ${\rm non-}D \bar D$ partial width
$\Gamma(\psi(3770) \rightarrow {\rm non-}D\bar D) = (2.8 \pm 1.8 \pm 2.4)$
MeV.
Meanwhile we determine the observed cross section for $non-D\bar D$
event production from $\psi(3770)$ decays to be
$\sigma^{\rm obs}(\psi(3770)\rightarrow non-D\bar D) =
(0.72 \pm 0.46 \pm 0.62)$ nb.
\end{abstract}

\maketitle

\section{Introduction}

The $\psi(3770)$ charmonium state is the lowest-mass resonance above open
charmed pair threshold~\cite{pdg04}. It is believed to mainly be
a mixture of the $1^3D_1$
and $2^3S_1$ angular momentum eigenstate. This charmonium state is expected
to predominately decay into the OZI allowed $D^0\bar D^0$ and $D^+D^-$
final states. 

However, there is a long-standing puzzle in the $\psi(3770)$ production
and decays. Available published data indicate that
the measured values of
$D \bar D$ production cross section at the peak of
$\psi(3770)$ resonance fail to fit in the measured values of the cross
section for $\psi(3770)$ production.
A detail analysis of the experimental data taken at SPEAR shows 
that about $38\%$ of $\psi(3770)$ does not decay into 
$D^0\bar D^0$ and $D^+D^-$ \cite{rgew04} (see section II for detail).
Some new publications about $\psi(3770)$
and $D \bar D$ production and decays from
BES~\cite{bes_xsct_ddbar}~\cite{bes_xsct_ddbar_dbltg}~\cite{bes_psipp_prmt} 
and CLEO~\cite{cleo_ddbar}
still indicate that $\psi(3770)$ is not saturated by $D^0\bar D^0$ and
$D^+D^-$ decays.

Measurements of the branching fractions for
$\psi(3770) \rightarrow D^0 \bar D^0, D^+D^-,D\bar D$
can help for understanding of the decay mechanism of $\psi(3770)$
and for elucidating the exact nature of the $\psi(3770)$.
Measurement of the partial width
for $\psi(3770) \rightarrow {\rm non-}D\bar D$ decay is essential 
to clarify the long-standing puzzle 
of the $\psi(3770)$ production and decays,
and is helpful for solving the $\rho-\pi$ puzzle in $\psi(3686)$
decay~\cite{jrosner_hep_ph_0105327}.

Recently, BES published a preliminary result on measurement of
the cross section for $\psi(3770)$ production obtained by analyzing
a fine cross section scan data sets taken from 3.666 GeV to 3.897 GeV.
Based on these modern measurements for both the $\psi(3770)$ and $D\bar D$
production at the peak of $\psi(3770)$ resonance and the measurement
of the $\psi(3770)$ resonance parameters and with making some necessary corrections 
for radiative effects, 
we can determine the branching fractions for
$\psi(3770) \rightarrow D^0 \bar D^0, D^+D^-$
and $\psi(3770) \rightarrow D\bar D$ in a consistent way 
with the new measured parameters by BES.
Meanwhile we can extract the branching fraction for 
$\psi(3770) \rightarrow {\rm non-}D\bar D$.

In this letter we report the determinations of the branching fractions.

\section{$\psi(3770)$ resonance parameters and cross section for $D\bar D$
production measured at SPEAR}

  There are several measurements of the $\psi(3770)$ resonance parameters.
Table~\ref{psipp} summarizes the measured values of the $\psi(3770)$
resonance parameters by the MARK-I~\cite{mark1}, 
DELCO~\cite{delco} and MARK-II~\cite{mark2} experiments at the SPEAR, and
the world average of the parameters given by Particle Data
Group~\cite{pdg04}.
\begin{table}
\caption{Comparison of the measured $\psi(3770)$ resonance parameters from
different experiments.}
\label{psipp}
\begin{tabular}{cccc} \hline \hline
Experiment  & $M$ (MeV) & $\Gamma_{tot}$ (MeV)  &
 $\Gamma_{ee}$ (eV)  \\
\hline
MARK-I~\cite{mark1}  &    3772$\pm$6     &   28$\pm$5   &  345$\pm$85  \\
DELCO~\cite{delco}   &    3770$\pm$6         &   24$\pm$5   &  180$\pm$60  \\
MARK-II~\cite{mark2} &    3764$\pm$5         &   23.5$\pm$5.0   &  276$\pm$50    \\ \hline
PDG2004~\cite{pdg04} &    3769.9 $\pm$ 2.5   & 23.6$\pm$2.7 &  260$\pm$40  \\ \hline
\hline
\end{tabular}
\end{table}

The production of $\psi(3770)$ resonance in the $e^+e^-$ annihilation can be
described by bare Breit-Wigner cross section
\begin{equation}
\sigma^{\rm bare}_{\psi(3770)}(E) = \frac{12 \pi \Gamma^0_{ee}\Gamma_{\rm tot}(E)}
{{(E^2-M^2)^2 + M^2\Gamma^2_{\rm tot}(E)}},
\end{equation}
where $\Gamma^0_{ee} = \Gamma_{ee}/(1+\delta_{\rm {\rm VP}})$,
$M$ and $\Gamma_{ee}$ are the mass and leptonic width
of the $\psi(3770)$ resonance respectively;
$E$ is the center-of-mass energy;
$\Gamma_{\rm tot}(E)$ is chosen to be energy dependent~\cite{bes_xsct_ddbar}.
At the peak of the $\psi(3770)$ resonance, the bare cross section for the
$\psi(3770)$ production is given by
\begin{equation}
\sigma^{\rm bare}_{\psi(3770)} = \frac{12 \pi}{M^2(1+\delta_{\rm VP})}
B(\psi(3770)\rightarrow e^+e^-)
\end{equation}
where the $B(\psi(3770)\rightarrow e^+e^-)$ is the branching fraction for
$\psi(3770)\rightarrow e^+e^-$ and $(1+\delta_{\rm VP})$ is 
a correction factor to correct the increased amount in the leptonic width 
due to the vacuum polarization effect including both leptonic and hadronic
terms. In the energy region from
charm threshold to 4.14 GeV, $(1+\delta_{VP})$ varies by
less than $\pm 2\%$~\cite{bes_xsct_ddbar}. 
In this section, we treat it as a constant of
\begin{equation}
    (1+\delta_{\rm VP})=1.047\pm0.024.
\end{equation}

Inserting the mass and the purely leptonic branching
fraction~\cite{pdg04} of $\psi(3770)$ into Eq. (2) we obtain the bare cross section for
$\psi(3770)$ production to be
\begin{equation}
\sigma^{\rm bare}_{\psi(3770)}|_{\rm peak} = (11.0 \pm 1.7)~~{\rm nb}
\end{equation}

MARK-III measured the observed cross section~\cite{mark3} for
$D\bar D$ production at center-of-mass energy $\sqrt{s}=3.768$ GeV to be
$\sigma^{obs}_{D\bar D}= 5.0 \pm 0.5~~ \rm{nb}$~\cite{mark3}.
At this energy, the cross section is reduced 
by a factor of about 0.734~\footnote{In the calculation of the bare cross
section, the $\psi(3770)$
resonance parameters
$M=3769.9 \pm 2.5$ MeV;
$\Gamma_{\rm tot}=23.6 \pm 2.7$ MeV and
$\Gamma_{ee} = 0.26 \pm 0.04$ keV
from Particle Data Group ~\cite{pdg04} were used.
} 
due to initial state radiation and photon vacuum porlarization effects
(see section V for more detail).
After correcting the observed cross section 
$\sigma^{obs}_{D\bar D}$ for the radiative effects,
we obtain the bare cross section for $D \bar D$ production to be
\begin{equation}
\sigma^B_{D \bar D} = 6.8 \pm 0.7~~\rm{nb}.
\end{equation}

The large discrepancy between $\sigma_{\psi(3770)}$ 
and $\sigma_{D \bar D}$
indicates that either, contrary to
what is generally expected,
$\psi(3770)$ could substantially decay into
non$-D\bar D$ final states
or the measured cross sections for 
$D\bar D$ and $\psi(3770)$ production
suffer from large systematic shifts. Otherwise, there might exist
some other effects which are responsible for the discrepancy.
If we simply calculate the branching fraction for 
$\psi(3770)\rightarrow non-D\bar D$ based on the above bare cross sections for
$\psi(3770)$ and $D\bar D$ production, we obtain
\begin{equation}
B(\psi(3770)\rightarrow non-D\bar D) = (38.2 \pm 11.5)\%.
\end{equation}
This value is obtained based on the different measurements at the SPEAR.
Howevere, people suspected 
that the large discrepancy could be due to systematic shift
in normalization, since no experiment has measured the
cross sections for both the $\psi(3770)$ and $D\bar D$ production
simultaneously.

The best way to solve this open question is to measure both the
bare cross sections for $\psi(3770)$ and $D\bar D$ production
from the same experiment.
The new measurements of the $\psi(3770)$ resonance parameters and
the cross sections for $D\bar D$ production from BES-II experiment
supply one an opportunity to clear the situation. 

\section{$\psi(3770)$ resonance parameters measured by BES-II}

BES Collaboration recently reported the results on measurements of
the $\psi(3770)$ resonance parameters~\cite{bes_psipp_prmt} 
with a higher precision.
Table~\ref{bes_psipp_psip_rslt}
summarizes the measured values~\cite{bes_psipp_prmt}, 
where the errors are the combined statistical
and systematic errors together. The paper~\cite{bes_psipp_prmt}
also reports the cross section for $\psi(3770)$ production at its peak to be
\begin{equation}
\sigma^{\rm prd}_{\psi(3770)}=9.07\pm 0.82~~~{\rm nb}.
\end{equation}

\noindent
This value of the cross section includes
the contribution from vacuum polarization effect.
After correcting the vacuum polarization
effect with the factor given by Eq. (3), we obtain the bare cross section 
for $\psi(3770)$ production at the peak of the resonance to be
\begin{equation}
\sigma^{\rm bare}_{\psi(3770)}=8.66\pm 0.81~~~{\rm nb},
\end{equation}
where the erorr includes the uncertainty in the correction 
to the vacuum polarization effect.

\begin{table}
\caption{The measured $\psi(3770)$ resonances parameters from
BES-II experiment.}
\label{bes_psipp_psip_rslt}
\begin{tabular}{cccc} \hline \hline
Resonance  & $M$ (MeV) & $\Gamma_{tot}$ (MeV)  & $\Gamma_{ee}$ (eV)  \\
\hline
$\psi(3770)$  &    3772.3$\pm$1.0     &   25.5$\pm$3.1   &  224$\pm$31  \\
\hline \hline
\end{tabular}
\end{table}

\begin{center}
\begin{figure}[hbt]
\includegraphics*[width=7.5cm,height=9.0cm]
{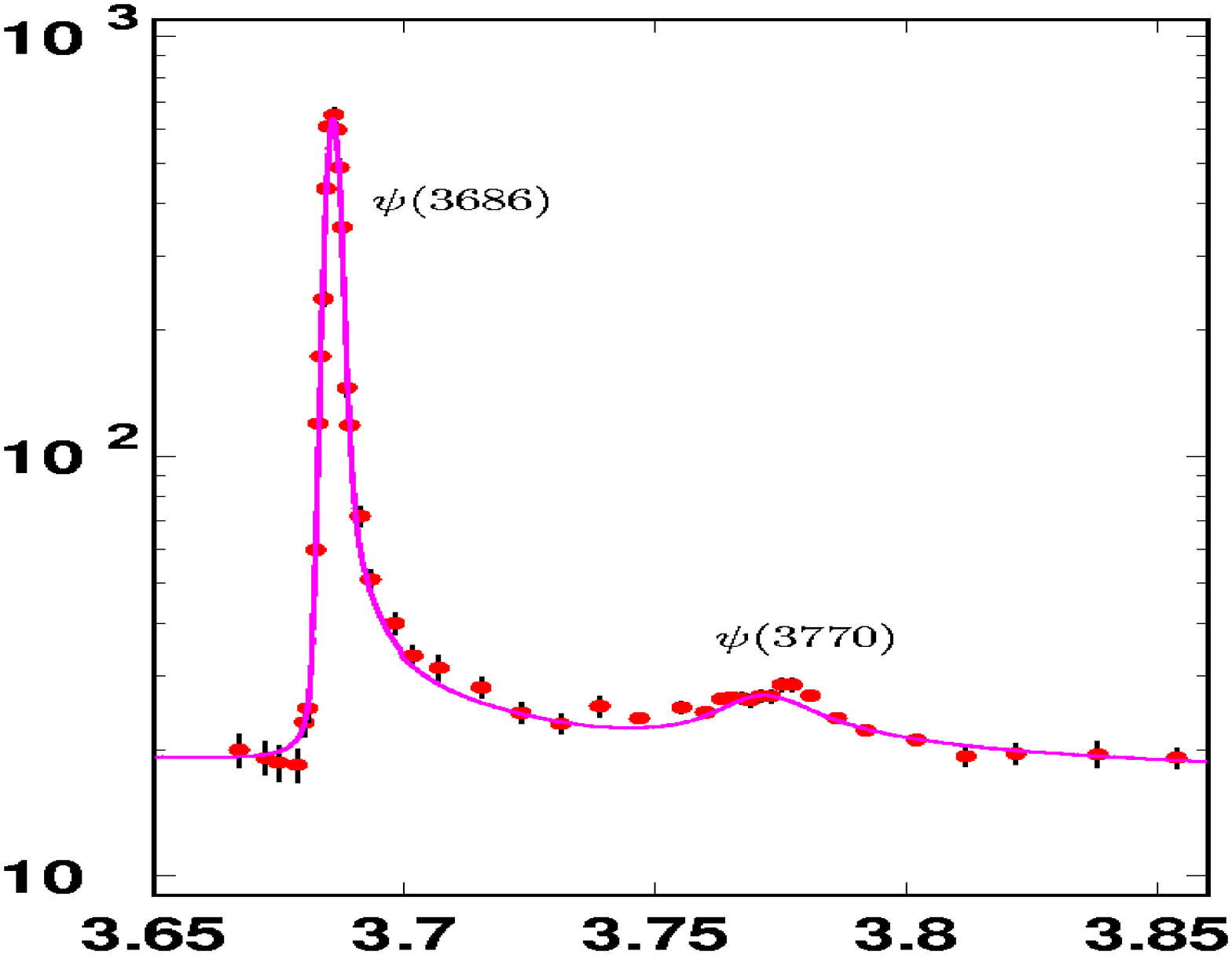}
\put(-125,-15){\bf\large $E_{\rm cm}$ [GeV]}
\put(-225,125){\rotatebox{90}{$\sigma^{\rm obs}_{\rm had}$~~~[nb]}}
\caption{The observed inclusive hadronic cross sections vs the nominal
c.m. energies; the error bars represent the observed cross sections and the
curve is the best fit which is discussed in more details in the
Ref.~\cite{bes_psipp_prmt}.
}
\label{xsct}
\end{figure}
\end{center}

\section{Cross sections for $D^0\bar D^0$, $D^+D^-$ and
$D\bar D$ production measured by BES-II experiment
at $\sqrt{s}=3.773$ GeV}

With single and double tag analysis method~\cite{bes_xsct_ddbar_dbltg},
BES Collaboration analyzed the invariant mass spectra of
$ K^-\pi^+$, $K^-\pi^+\pi^0$, $K^-\pi^+\pi^+\pi^-$, $K^- \pi^+\pi^+$,
$K^0_S \pi^+$, $K^0_S \pi^+\pi^+\pi^-$, $K^- K^+\pi^+$,
$K^- K^+\pi^+$, $K^0_S K^+$, $K^-\pi^+\pi^+\pi^0$ and
$K^0_S \pi^+\pi^0$ combinations for single tag analysis, 
and the mass spectra of $K^-\pi^+$, $K^-\pi^+\pi^0$,
$K^-\pi^+\pi^+\pi^-$ and 
$K^-\pi^+\pi^+$ combinations for double tag analysis
to measure the observed cross sections for $D^0\bar D^0$ and
$D^+D^-$ production at $\sqrt{s}=3.773$ GeV. They obtained the cross sections
for $D^0 {\bar D}^0$, $D^+D^-$ and $D\bar D$ production
at this energy point to be~\cite{bes_xsct_ddbar_dbltg}
\begin{equation}
\sigma_{D^0 \bar {D^0} }^{\rm obs} = (3.47 \pm 0.32 \pm 0.21)~{\rm nb},
\end{equation}
\begin{equation}
     \sigma_{D^+  {D^-} }^{\rm obs} = (2.46 \pm 0.33 \pm 0.20)~~ {\rm nb},
\end{equation}
\begin{equation}
   \sigma_{D \bar {D} }^{\rm obs} = (5.93 \pm 0.46 \pm 0.35)~~ {\rm nb},
\end{equation}
where the first error is statistical and second systematic. 
These values of the cross sections are consistent with that measured by BES
Collaboration based on single tag analysis~\cite{bes_xsct_ddbar},
and consistent with that measured by CLEO~\cite{cleo_ddbar}
within error.

To obtain the bare cross sections for $D^0\bar {D^0}$, $D^+D^-$ and
$D\bar D$ production, the observed cross sections 
have to be corrected for the radiative effects.

\section{Radiative Corrections}
In any $e^+e^-$ colliding beam experiment, the electron (positron) always
radiates at the interaction point because of the potential of the
positron (electron).
Since this radiation (Bremsstrahlung) carries
energy away,
the actual center-of-mass energy for the $e^+e^-$ annihilation is reduced by
Bremsstrahlung to $\sqrt{s(1-x)}$, where $xE_{\rm beam}$ is the total energy
of the emitted photons.
The Bremsstrahlung is principally responsible for the distortions
to the bare resonance line shape, while the self energy
of the electron and positron and the vertex corrections to the initial state
affect the overall factors to change the scale of the cross section.
All of these corrections are called Initial State Radiation (ISR)
corrections.
The bare cross section for $D \bar D$ production at the energy
of 3.773 GeV can be obtained by correcting the
observed cross section for the effects of the ISR and
vacuum polarization.

The observed cross section, $\sigma^{\rm obs}$,
at the nominal energy $\sqrt{s}$ can be written as a convolution of the
bare cross section $\sigma^B(s(1-x))$ and
a sampling function $F(x,s)$~\cite{kuraev}~\cite{bes_xsct_ddbar},
\begin{equation}
  \sigma^{\rm obs}(s)= \int^1_0 dx
             F(x,s) \frac{\sigma^{B}(s(1-x))}{|1-\Pi(s(1-x))|^2 },
\end{equation}
\noindent
in which the $|1-\Pi(s(1-x))|^{-2}$ is due to the photon vacuum
polarization,
the $\Pi(s)$ is the photon vacuum polarization operator
including the loops over leptons ($e$ $\mu$ and $\tau$)
and hadrons~\cite{berends}~\cite{andrej}.

Since we are interested in the $\psi(3770)$ resonance
in this analysis, we take the $\sigma^B$ to be the bare
Breit-Wigner cross section as given in Eq. (1),
where
$\Gamma_{\rm tot}(E)$ is chosen to be energy dependent and normalized to
the total width $\Gamma_{\rm tot}$
at the peak of the resonance~\cite{pdg04}~\cite{mark2}.
The $\Gamma_{\rm tot}(E)$ is defined as~\cite{bes_xsct_ddbar}~
\begin{equation}
 \Gamma_{\rm tot}(E) =  \Gamma_{\rm tot}  \frac{  \frac{p^3_{D^0}}{1+(r
{p_{D^0}})^2}
                          + \frac{p^3_{D^{\pm}}}{1+(r {p_{D^{\pm}}})^2} }
                 {  \frac{{p^0}^3_{D^0}}{1+(r {{p^0}_{D^0}})^2}
                          + \frac{{p^0}^3_{D^{\pm}}}{1+(r
{{p^0}_{D^{\pm}}})^2} },
\end{equation}
where $p^0_D$ is the momentum of the $D$ mesons produced at the
peak of $\psi(3770)$, $p^{ }_D$ is the momentum of the $D$ mesons
produced at the c.m. energy $\sqrt{s}$, $\Gamma_{\rm tot}$ is the width of
the $\psi(3770)$ at the peak,
and $r$ is the interaction radius which was set to be 0.5 fm in the analysis.
In the calculation of the bare cross section, the $\psi(3770)$
resonance parameters 
$M=3772.3 \pm 1.0$ MeV;
$\Gamma_{\rm tot}=25.5 \pm 3.1$ MeV and
$\Gamma_{ee} = 0.224 \pm 0.031$ keV
measured by BES Collaboration ~\cite{bes_psipp_prmt} were used.

  The $\psi(3770)$ width ($\sim 25$ MeV) is much large than the energy
spread ($\sim 1.37 $ MeV) of the BEPC. So
the effect of the beam energy spread
on the cross section could be ignored. The $\psi(3770)$
is generally assumed to decay exclusively into $D \bar D$.
Taking these considerations,
the observed cross section of Eq. (12)
should be replaced by
\begin{equation}
  \sigma^{\rm obs}(s)= \int^{1-4M_D^2/s}_0  dx
F(x,s) \frac{\sigma^{B}(s(1-x))}{|1-\Pi(s(1-x))|^2 } 
\end{equation}
\noindent
in calculation of the radiative corrections.
The correction factor for the radiative effects is given by
\begin{equation}
 g = \frac{\sigma^{\rm obs}}{\sigma^{B}}.
\end{equation}
\noindent
Figure 2 shows the factor of the radiative corrections as a function
of the nominal center-of-mass energy. At the center-of-mass energy
$\sqrt{s}=3.773~ \rm GeV$, the factor is
\begin{equation}
 g = 0.768 \pm 0.020,
\end{equation}
\noindent
where the error is the uncertainty arising from the errors of the
$\psi(3770)$ resonance parameters and the uncertainty 
in vacuum polarization correction. 
\begin{figure}
\includegraphics[width=11.0cm,height=9.0cm]
{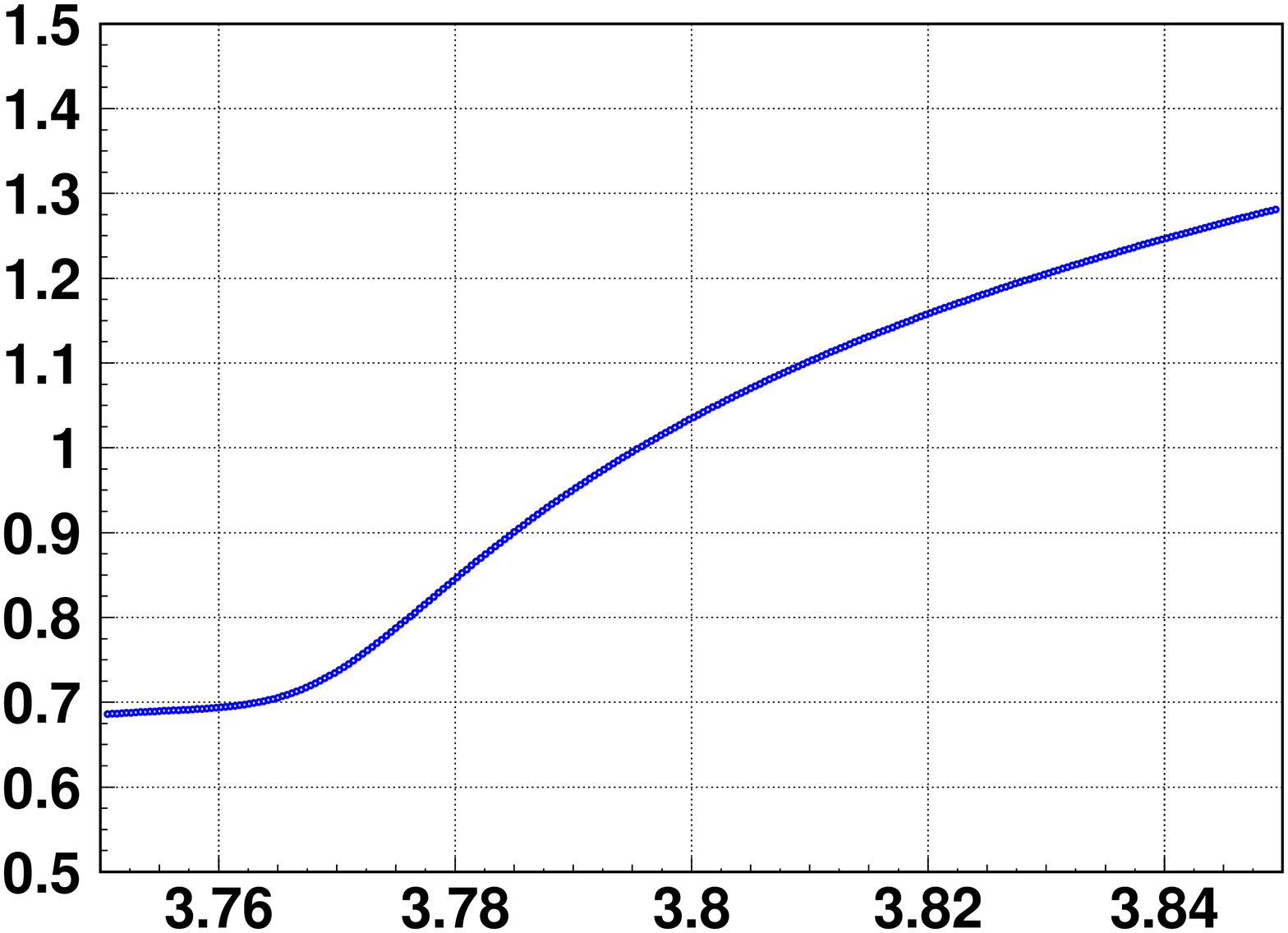}
\put(-190.0,-15.0){\bf\large {Nominal $E_{cm}$~~~~[GeV]}}
\put(-315.0,170.0){\Large{$g$}}
\caption{The factor of radiative corrections as a function
of the nominal center-of-mass energy.}
\end{figure}

\section{bare cross section for $D \bar D$ production}

    The bare cross sections for $D^0 \bar D^0$, $D^+D^-$ and  $D\bar D$ 
production are obtained by dividing the observed cross sections by the factor
$g=0.768\pm0.020$
of the radiative corrections. At $\sqrt{s}=3.773$ GeV,
the bare cross sections are
\begin{equation}
    \sigma^{\rm bare}_{D^0 {\bar D}^0} = (4.52 \pm 0.42 \pm 0.30)~~\rm nb,
\end{equation}
\begin{equation}
    \sigma^{\rm bare}_{D^+D^-} = (3.20 \pm 0.43 \pm 0.27)~~\rm nb,
\end{equation}
\noindent
and
\begin{equation}
    \sigma^{\rm bare}_{D \bar D} = (7.72 \pm 0.60 \pm 0.50)~~\rm nb,
\end{equation}
\noindent
where the first error is statistical and the second systematic which
include the uncertainty in the factor 
of the radiative corrections ($\sim 2.6\%$).

\section{Branching fractions for $\psi(3770)\rightarrow D\bar D$
and $\psi(3770)\rightarrow non-D\bar D$}

Dividing the bare cross sections for $D^0\bar D^0$, $D^+D^-$ and $D\bar D$
production by the bare cross section for $\psi(3770)$ prodcution 
given by Eq. (8), we obtain
the branching fractions to be
\begin{equation}
B(\psi(3770) \rightarrow D^0\bar D^0)=(52.2 \pm 4.8 \pm 5.5)\%,
\end{equation}
\begin{equation}
B(\psi(3770) \rightarrow D^+D^-)=(37.0 \pm 5.0 \pm 4.3)\%,
\end{equation}
\begin{equation}
B(\psi(3770) \rightarrow D\bar D)=(89.1 \pm 6.9 \pm 9.2)\%
\end{equation}
\noindent and the latter one implies
\begin{equation}
B(\psi(3770) \rightarrow non-D\bar D)=(10.9 \pm 6.9 \pm 9.2)\%,
\end{equation}
\noindent
where the first error is statistical and second the uncancelled
systematic uncertainty. The systematic uncertainties in luminosity
measurement and in the radiative correction factor are canceled 
in the determinations of the branching fractions.
Using the total width of $\psi(3770)$ resonance 
$\Gamma^{\psi(3770)}_{\rm tot}= 25.5 \pm 3.1$ MeV measured by the BES
Collaboration, this branching fraction corresponds to a partial width of
\begin{equation}
\Gamma(\psi(3770) \rightarrow non-D\bar D)=
(2.8  \pm 1.8 \pm 2.4)~~~{\rm MeV},
\end{equation}
where the first error is statistical and second systematic arising from the
uncertainty in the measured branching fraction for 
$\psi(3770) \rightarrow {\rm non}-D\bar D$ and the uncertainty in the
measured total width of $\psi(3770)$ resonance.
Those result in a total observed cross section for ${\rm non}-D\bar D$ final
states of $\psi(3770)$ decays at $\sqrt{s}=3.773$ GeV to be
\begin{equation}
\sigma^{\rm obs}(\psi(3770) \rightarrow {\rm non}-D\bar D) = 
(0.72 \pm 0.46 \pm 0.62)~~\rm nb,
\end{equation}
where the first error reflects the statistical fluctuation on the measured
branching fraction for the decay $\psi(3770) \rightarrow {\rm non}-D\bar D$
and the second one reflects all of other uncertainties, such as the systematic
uncertainty in the measured branching fraction, 
the uncertainty in the radiative effect corrections
and the uncertainty in the total width of $\psi(3770)$ resonance.

\section{Summary and discussion}

Based on the recent publications about $\psi(3770)$
and $D\bar D$ production and decays
from BES Collaboration and 
with making some necessary corrections 
for initial state radiative and
vacuum polarization effects, 
we determined the branching fractions for
$\psi(3770) \rightarrow D^0 \bar D^0, D^+D^-, D\bar D$ to be
$(52.2 \pm 4.8 \pm 5.5)\%$, $(37.0 \pm 5.0 \pm 4.3)\%$,
and $(89.1 \pm 6.9 \pm 9.2)\%$, respectively.
Meanwhile we determined the branching fraction for
$\psi(3770) \rightarrow {\rm non-}D\bar D$ to be
$(10.9 \pm 6.9 \pm 9.2)\%$,
corresponding the partial width 
$\Gamma(\psi(3770) \rightarrow {\rm non-}D\bar D) = 
(2.8 \pm 1.8 \pm 2.4)$ MeV.
We also determined the observed cross section for
$\psi(3770) \rightarrow {\rm non-}D\bar D$ to be
$\sigma^{\rm obs}(\psi(3770) \rightarrow {\rm non-}D\bar D)=
(0.72 \pm 0.46 \pm 0.62)$ nb
at $\sqrt{s}=3.773$ GeV.

There previously were some estimations 
on the branching fraction for the decay 
$\psi(3770)\rightarrow non-D\bar D$~\cite{rgew04}~\cite{jrosner_hep_ph_0405196}. 
These estimations were based on some measurements 
of the cross sections for $\psi(3770)$ and
$D\bar D$ production from different 
experiments~\cite{mark1}~\cite{delco}~\cite{mark2}~\cite{crystal_ball}~~\cite{bes2_r}. 
In that case there
may be some systematic shifts in the measured values of
$\sigma_{\psi(3770)}$ and $\sigma_{D\bar D}$ 
due to different normalizations with luminosities of data sets 
and Monte Carlo efficiencies. 
However, all of the physics quantities determined in this letter
are based on the measurements from the same experiment, and
the cross sections for both the $\psi(3770)$ resonance and $D\bar D$ 
production are transfered into the bare ones 
by making the necessary corrections for the initial
state radiative and vacuum polarization effects.  In this case
the results of the determined physics quantities would be more reliable. 

We are expecting 
that both BES and CLEO Collaborations will give more precise
measurements of the partial width for the decay 
$\psi(3770) \rightarrow {\rm non-}D\bar D$ in the near future.

\end{document}